\documentclass[twocolumn,prb]{revtex4}% Physical Review B

\usepackage{graphicx}% Include figure files
\usepackage{dcolumn}% Align table columns on decimal point
\usepackage{bm}% bold math

\begin{document}

\preprint{submitted to PRB}

\title{Anomalous broadening of the spin-flop transition in the reentrant spin-glass phase of La$_{2-x}$Sr$_x$CuO$_4$ ($x=0.018$)}

\author{T. Suzuki, and T. Goto}
\address{Institute for Materials Research, Tohoku University, Sendai 980-8577, Japan}
\author{K. Chiba, and T. Fukase}
\address{Institute for Materials Research, Tohoku University, Sendai 980-8577, Japan}
\author{M. Fujita, and K. Yamada}
\address{Institute for Chemical Research, Kyoto University, Uji Kyoto, 610-0011, Japan}

\date{\today}% It is always \today, today,

\begin{abstract}
The magnetization in a lightly doped La$_{2-x}$Sr$_x$CuO$_4$ ($x=0.018$) single crystal was measured.
Spin-flop transition was clearly observed in the hole doped antiferromagnetically ordered state under increasing magnetic fields perpendicular to the CuO$_2$ plane.
In the spin-glass phase below 25K, the spin-flop transition becomes broad but the step in the magnetization curve associated with the transition remains finite at the lowest temperature.
We show in this report that, at low temperature, the homogeneous antiferromagnetic order is disturbed by the re-distribution of holes, and that the spatial variance of the local hole concentration around $x=0.018$ increases.
\end{abstract}

\pacs{74.72.Dn}

\maketitle

\indent
Magnetic properties of lightly doped La$_{2-x}$Sr$_x$CuO$_4$ have been investigated intensively because of the strong interest in the relation between the two-dimensional Heisenberg antiferromagnet and the mechanism of the high-{\it T}$_{\rm c}$ superconductivity.
The end-member material La$_2$CuO$_4$ shows an aniferromagnetic transition at {\it T}$_{\rm N}=$320 K, and its ordered spins lie almost in the CuO$_2$ plane~\cite{Yama,Vaknin}.
In La$_{2-x}$Sr$_x$CuO$_4$ (LSCO), the three-dimensional (3D) antiferromagnetic order disappears rapidly with hole doping and a spin-glass state appears in the region of $0.02<x< 0.06$~\cite{Harshman,Cho1,Hayden,Kastner,Keimer,Aharony,Chou,Yamada}.
Remarkable progress in understanding the character of the spin-glass state was brought about by the magnetization and the neutron scattering measurements of single crystals~\cite{Wakimoto1,Wakimoto2,Matsuda1,Matsuda2,Ando}.
Wakimoto {\it et al}. and Matsuda {\it et al}. confirmed the diagonal spin modulation in $0.03<x<0.05$ by elastic neutron scattering measurements.
In the diagonal stripe structure, the positions of the incommensurate magnetic peaks are rotated 45$^{\circ}$ in reciprocal space around the ($\pi$,$\pi$) position compared to those of the superconducting phase.
For a concentration of $x=0.02$ at which the N\'eel temperature goes to zero, a cluster model, {\it i}.{\it e}., a random freezing of quasi-three-dimensional antiferromagnetically ordered spin clusters has been proposed to explain the spin-glass like behavior~\cite{Matsuda1}.
In short, the magnetic properties at a Sr-concentration of $0.02<x<0.06$ have been investigated rather well.
However, in the range of $0<x<0.02$ inside the antiferromagnetically ordered region of the phase diagram, magnetic properties are still unknown except for a few investigations by NQR and $\mu$SR measurements using polycrystalline samples~\cite{Chou2,Borsa,Nied}.
In this study, we report on magnetization measurements in a La$_{2-x}$Sr$_x$CuO$_4$ ($x=0.018$) single crystal, with $x$ slightly smaller than the disappearance point of the 3D-antiferromagnetic order with hole doping. 
\vspace{11pt}
\\
\indent
The single crystal used in this study was grown by the traveling-solvent-floating-zone (TSFZ) method in a gas consisting of 90\%-argon and 10\%-oxygen at 1 atm.
The sample size is $\sim$3$\times$3$\times$5mm.
In order to eliminate excess oxygen, the sample was annealed at 1 bar of argon gas flowing at 900 $^{\circ}$C for 24 hours.
The temperature dependence of the magnetization was measured at applied fields of  0.1 T and 1 T using a Quantum Design SQUID magnetometer.
Sweeping field measurements between 0 T and 8 T ({\it M}-{\it H} curve) were performed with an Oxford's vibrating sample magnetometer. 
The sweep rate was $\sim$0.25 Tesla/min. and the vibration frequency was 45 Hz.
The direction of the magnetic field was perpendicular to the CuO$_2$ plane in all measurements.
\vspace{11pt}
\\
\indent
Figure 1 shows the temperature dependence of the magnetic susceptibility at fields of 0.1 T and 1 T measured once under field cooling (FC) and once warming up after zero-field cooling (ZFC).
The magnetization increases with decreasing temperatures and shows a kink at 140K (inset of fig.1), corresponding to the antiferromagnetic order transition temperature {\it T}$_{\rm N}$.
The arrow in the figure indicates {\it T}$_{\rm N}$ as determined by neutron scattering~\cite{Matsuda3}.
The temperature dependence of the magnetization in the ZFC measurement begins to deviate from that of the FC measurement below 40K at 0.1 T, and below 10 K at 1 T.
It seems that the deviation of the magnetization in the ZFC measurement is suppressed by the magnetic field.
\begin{figure}[hhh]
\includegraphics[40mm,30mm][62mm,116mm]{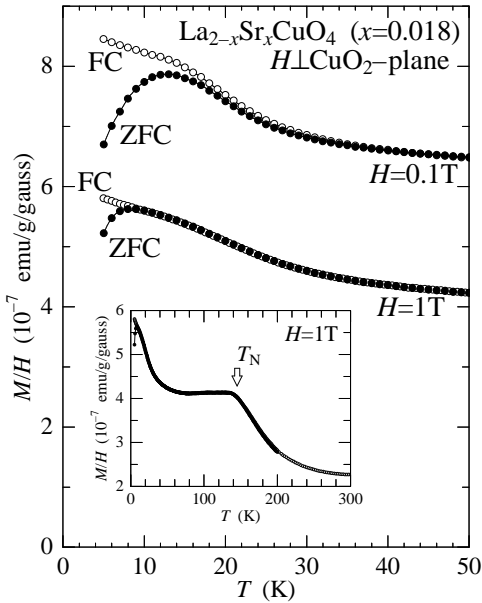}
\caption{\label{fig:epsart} The temperature dependence of the magnetic susceptibility of La$_{2-x}$Sr$_x$CuO$_4$ ($x=0.018$) in 0.1 T and 1 T, measured under field cooling (FC) and under warming up after zero-field cooling (ZFC).
Open circles and closed circles represent the FC and the ZFC data.
The magnetic field is applied perpendicularly to the CuO$_2$-plane.
The inset shows the data from 4.2 K to 300 K in case of {\it H} = 1 T.
The arrow indicates the {\it T}$_{\rm N}$ as determined by neutron scattering(ref.19).}
\end{figure}
These differences between the FC and ZFC measurements indicate a change in the spin system from the 3D-antiferromagnetically ordered state to a spin-glass phase at low temperatures.
\vspace{11pt}
\\
\indent
In the lightly doped LSCO, antiferromagnetically ordered spins cant from the CuO$_2$ plane due to the Dzyaloshinskii-Moriya (DM) interaction arisen from the orthorhombic distortion~\cite{Dzy,Mori,Thio1,Thio2}.
In zero-fields, the weak ferromagnetic moment of canted spins cancels out because the canting direction is opposite in alternate layers, but under increasing magnetic fields  perpendicular to the CuO$_2$ plane, the spins in all layers cant in the same direction along the applied magnetic field, and a spin-flop transition takes place.
As a result, a macroscopic weak ferromagnetic moment appears in the antiferromagnetically ordered state.
Thus, this spin-flop transition is considered to be a field-induced weak ferromagntic transition.
In case of La$_2$CuO$_4$, the spin-flop transition point {\it H}$_{\rm sf}$ is 5.3 T at the lowest temperature~\cite{Thio1}.
This transition is sensitive to changes in the magnitude of the weak ferromagnetic moment and can be utilized for the investigation of the ordered state.
\\
\indent
Figure 2 shows typical {\it M}-{\it H} curves at several temperatures.
All measurements were carried out with increasing and decreasing magnetic fields after ZFC conditions for each temperature.
A step in the {\it M}-{\it H} curve associated with the weak ferromagnetic moment appears below 140 K.
This temperature coincides with the known N\'eel temperature {\it T}$_{\rm N}$ in case of $x=0.018$.
This means that the impurity phase La$_2$CuO$_4$ does not exist in the used sample.
\begin{figure}[hhh]
\includegraphics[40mm,22mm][62mm,125mm]{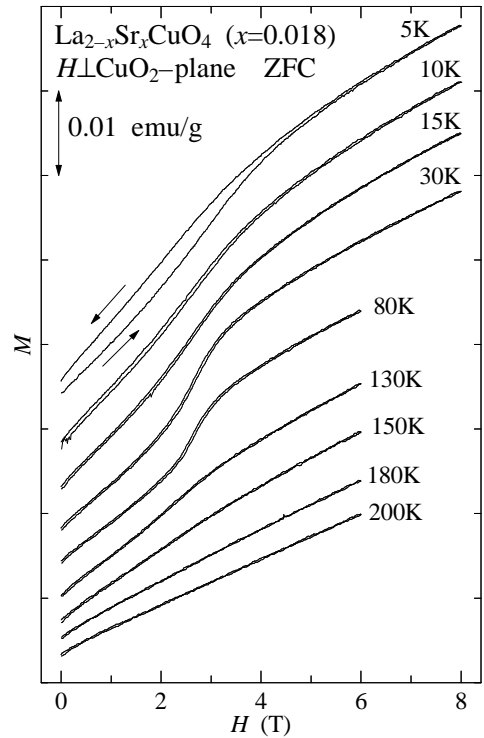}
\caption{\label{fig:epsart} Typical magnetic field dependence of the magnetization ({\it M}-{\it H} curve) at several temperatures.
All measurement were carried out with increasing and decreasing magnetic fields after zero-field cooling.}
\end{figure}
If La$_2$CuO$_4$ contributes to the magnetization, the step in {\it M}-{\it H} curves must be observed above 140K because the N\'eel temperature of the stoichiometric or a slightly doped La$_2$CuO$_4$ is around 300K.
Actually, the spin-flop transition occurs only below {\it T}$_{\rm N}=$140K.
The observed weak ferromagnetic transition is that appeared in the hole doped 3D-antiferromagnetically ordered state.
As is clear from fig.2, this transition is sharp above 25K, and at low temperatures, the significant broadening is observed though the finite step height is remained.
We analyze the temperature dependence of the {\it M}-{\it H} curve and discuss the drastic broadening of the spin-flop transition at low temperatures.
\\
\indent
Figure 3(a) shows the magnetic field dependence d{\it M}/d{\it H} at several temperatures.
Figure 3(b) shows the temperature dependence of the spin-flop transition point {\it H}$_{\rm sf}$ and the transition width $\Delta${\it H}.
We define the transition point {\it H}$_{\rm sf}$ as the position of the peak and the transition width $\Delta${\it H} as the full width at half-maximum (FWHM) of the d{\it M}/d{\it H} curves.
{\it H}$_{\rm sf}$ shows no significant change throughout the whole temperature range below 140 K.
The previously reported {\it H}$_{\rm sf}$ of La$_2$CuO$_4$ is $\sim$5.3 T at the lowest temperature and is also nearly constant against {\it T}~\cite{Thio1}.
Thus, {\it H}$_{\rm sf}$ depends on the hole concentration and not on temperature.
\begin{figure}[hhh]
\includegraphics[40mm,25mm][63mm,127mm]{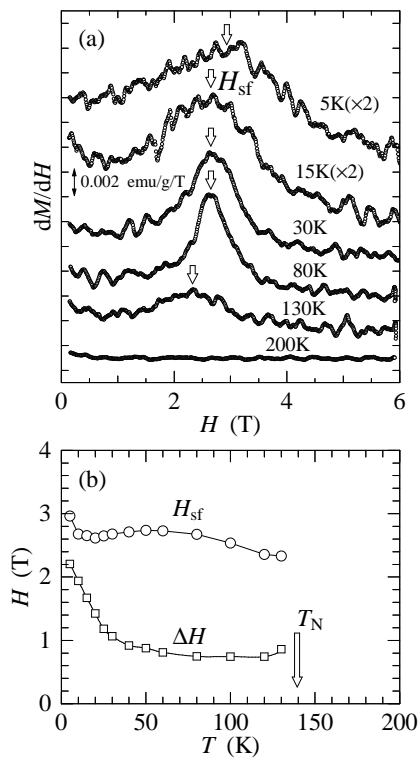}
\caption{\label{fig:epsart} (a) Magnetic field dependence of d{\it M}/d{\it H} at several temperatures.
Open arrows indicate the spin-flop transition point {\it H}$_{\rm sf}$.
(b)The temperature dependence of the spin-flop transition point {\it H}$_{\rm sf}$ (open circles) and the transition width $\Delta${\it H} (open squares) of ZFC measurements.}
\end{figure}
As evident from Fig.3(a), the transition width is narrow and the transition point is completely separated from {\it H}$_{\rm sf}$ of La$_2$CuO$_4$ above 25 K.
From these results, we conclude that, above 25 K, the hole distribution in the CuO$_2$ plane is homogeneous, {\it i}.{\it e}., the spatial variance of the hole concentration around the mean value $x=0.018$ is very small.
\vspace{11pt}
\\
\indent
Figure 4 shows the temperature dependence of $\Delta${\it M}.
{\it M}-{\it H} curve in the inset of fig.4 indicates the definition of the step height $\Delta${\it M}, from which we can deduce the weak ferromagnetic moment {\it M}$_{\rm F}$ to be $\sim1.7\times 10^{-3}\mu_{\rm B}$ per Cu atom at 25 K, about 80\% of that of La$_2$CuO$_4$~\cite{Thio1}.
\begin{figure}[hhh]
\includegraphics[40mm,35mm][63mm,89mm]{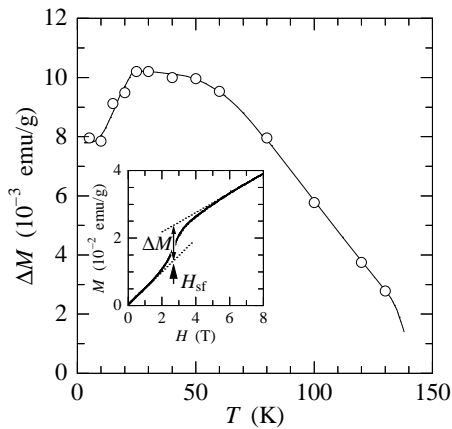}
\caption{\label{fig:epsart} Temperature dependence of the step-height of the magnetization $\Delta${\it M}.
The solid line is a guide for the eye.
The inset shows the definition of the step-height of the magnetization $\Delta${\it M} in the spin-flop transition.}
\end{figure}
These two comparable values suggest that nearly all Cu-3{\it d} spins contribute to the spin-flop above 25 K.
Below 25 K, $\Delta${\it M} drops to 80\% of that at 25 K.
That implies the two important facts.
1) The number of spins participating in the spin-flop decreases below 25 K because of the onset of the spin-glass phase.
2) Even below 25K, there does exist a number of antiferromagnetically ordered spins, because a spin-flop transition could not be caused in an assemblage of randomly directed spins.
Therefore, the spin-glass state observed below 25 K precipitates on the antifferomagnetically ordered state.
\vspace{11pt}
\\
\indent
Next, we discuss the residual antiferromagnetically ordered spins below 25 K participating in the spin-flop.
As shown in Fig.3, the spin-flop transition becomes anomalously broad below 25 K, and the peak of d{\it M}/d{\it H} curve spreads over 1$\sim$5 T at 5 K.
This means an increase of the spatial variance of the local hole concentration, {\it i}.{\it e}., a re-distribution of holes takes place below 25 K.
In consequence, spins in high local hole concentrations participate in  the spin-glass, while those in low hole concentrations contribute to the spin-flop.
\vspace{11pt}
\\
\indent
Finally, we discuss recent reports on elastic neutron scattering measurements in lightly-doped La$_{2-x}$Sr$_x$CuO$_4$ ($0<x<0.02$) under zero-field~\cite{Matsuda3} and high magnetic fields conditions~\cite{Matsuda4}.
They observed magnetic Bragg peaks and incommensurate magnetic peaks simultaneously at low temperatures at zero-field.
At decreasing temperatures, the intensity of the (100) magnetic Bragg peak decreases below 25 K and approaches a finite value at low temperatures.
Since the elastic neutron scattering intensity~\cite{Matsuda3} is proportional to the square of the ordered moment, we can roughly estimate from these data that the total ordered moment at low temperature is reduced to $\sim$70\% compared to that at 25 K.
The residual ratio of the total ordered moment as estimated from the temperature dependence of $\Delta${\it M} is $\sim$80\% (fig.4) under the plausible assumption that the antiferromagnetically ordered moment is in proportion to the weak ferromagnetic moment.
Thus the two observations are consistent.
At high fields, the spin-flop is detected and the broadening of the transition at low temperature is observed also by elastic neutron scattering~\cite{Matsuda4}.
They introduced an idea of the spin-cluster model that the long-range antiferromagnetically ordered region changes into assembled spin clusters and that the cluster size is finite and distributed.
Their cluster model is consistent with our results if we take into account the strong possibility that assembled spin clusters appear when the re-distribution of holes occurs.
\vspace{11pt}
\\
\indent
In summary, the magnetization of the single crystal La$_{2-x}$Sr$_x$CuO$_4$ ($x=0.018$) was measured.
The spin-flop transition is observed below {\it T}$_{\rm N}$=140 K, and the sharpness of this transition indicates that the doped holes are homogeneously distributed in space above 25 K.
In the spin-glass phase below 25 K, the character of the spin-flop transition changes drastically, {\it i}.{\it e}., the transition becomes broad and the step-height $\Delta${\it M} in the {\it M}-{\it H} curve associated with the weak ferromagnetic moment decreases, towards a finite value at low temperatures.
Thus, below 25K, the re-distribution of holes takes place, and the spatial variance of the local hole concentration around $x=0.018$ increases.
In consequence, the spins in the high local hole concentrations show the spin glass behavior while those in the low hole concentration contribute to the spin-flop.
\vspace{11pt}
\\
\indent
All the measurements of the magnetization were carried out at the Center for Low Temperature Science, Tohoku University.
The authors are grateful to Dr. M. Matsuda for useful discussion and Professor L. Boesten for a critical reading of the manuscript.
This research was partially supported by the Ministry of Education, Science, Sports and Culture, Grant-in-Aid for Scientific Research on Priority Areas (Novel Quantum Phenomena in Transition Metal Oxides), 12046256, 2000 and 12046239, 2001; for Scientific Research (C), 12040360, 2000; for Scientific Research (A), 10304026, 2001; Grant-in-Aid for Encouragement of Young Scientists, 13740216, 2001, and Grant-in-Aid for Creative Scientific Research (13NP0201) "Collaboratory on Electron Correlations - Toward a New Research Network between Physics and Chemistry -", and was partially supported by the Japan Science and Technology Corporation, under the Core Research for Evolutional Science and Technology Project (CREST).

\end{document}